\documentstyle[aps,preprint]{revtex}

\begin{document}
\draft
\title{Thermal fluctuations and disorder effects in vortex lattices}
\author{Dingping Li\thanks{%
e-mail:lidp@phys.nthu.edu.tw}}
\address{{\it National Center for Theoretical Sciences} \\
{\it P.O.Box 2-131, Hsinchu, Taiwan, R. O. C.}}
\author{Baruch Rosenstein\thanks{%
e-mail: baruch@phys.nthu.edu.tw}}
\address{{\it National Center for Theoretical Sciences and} \\
{\it Electrophysics Department, National Chiao Tung University } \\
{\it Hsinchu 30050, Taiwan, R. O. C.}}
\date{\today}
\maketitle

\begin{abstract}
We calculate using loop expansion the effect of fluctuations on the
structure \ function and magnetization of the vortex lattice and compare it
with existing MC results. In addition to renormalization of the height of
the Bragg peaks of the structure function, there appears a characteristic
saddle shape ''halos'' around the peaks. The effect of disorder on
magnetization is also calculated. All the infrared divergencies related to
soft shear cancel.
\end{abstract}


\newpage

\section{Introduction}

Decoration \cite{Gammel}, neutron scattering \cite{Keimer} and STM \cite
{Maggio} clearly demonstrated the Abrikosov flux line lattice in low and
high $T_{c}$ type II superconductors. There are however important
differences between the two classes of materials. Ginzburg parameter $Gi$
characterizing importance of thermal fluctuations is much larger in high $%
T_{c}$ superconductors than in the low temperature ones. Moreover in the
presence of magnetic field the importance of fluctuations in high $T_{c}$
superconductors is further enhanced. The lattice melts and becomes vortex
liquid over large portions of the phase diagram \cite{Zeldov,Nelson,Blatter}%
. In ''strongly fluctuating'' superconductors, even far below the melting
line, corrections to various physical quantities like magnetization or
specific heat are not negligible. The vortex lattice becomes distorted. It
is quite straightforward to systematically account for the fluctuations
effect on magnetization, specific heat or conductivity perturbatively above
the mean field transition line using Ginzburg - Landau (GL)\ description 
\cite{Tinkham}. However in the interesting region below this line it turned
out to be extremely difficult to develop a quantitative theory.

A direct approach to the low temperature fluctuations physics is to start
from the mean field solution and then take fluctuations around this
inhomogeneous solution into account perturbatively. Experimentally it is
reasonable since, for example, specific heat at low temperatures is a smooth
function and the fluctuations contribution is quite small. For some time
this was in disagreement with theoretical expectations. Eilenberger
calculated spectrum of harmonic excitations of the triangular vortex lattice 
\cite{Eilenberger} and noted that the gapless mode is softer than the usual
Goldstone mode expected as a result of spontaneous breaking of translational
invariance. The inverse propagator for the ''phase'' excitations behaves as $%
k_{z}^{2}+const(k_{x}^{2}+k_{y}^{2})^{2}$. It was shown \cite{Ikeda,Moore}
that this behavior is directly related to the nondispersive nature of the
shear modulus $c_{66}$ and is in agreement with numerous experiments.\ 

The influence of this additional ''softness'' goes beyond enhancement of the
contribution of fluctuations at leading order. It apparently leads to
disastrous infrared divergencies at higher orders rendering the perturbation
theory around the vortex state doubtful. One therefore tends to think that
nonperturbative effects are so important that such a perturbation theory
should be abandoned \cite{Ruggeri}. However it was shown in \cite{Rosenstein}
that a closer look at the diagrams reveals that in fact one encounters
actually only logarithmic divergencies. This makes the divergencies similar
to so called ''spurious'' divergencies in the theory of critical phenomena
with broken continuous symmetry and they exactly cancel at each order
provided we are calculating a symmetric quantity. One can effectively use
properly modified perturbation theory to quantitatively study various
properties of the vortex liquid phase. Magnetization calculated using this
perturbative approach agrees very well with the direct Monte Carlo
simulation of \cite{Sasik}. The method was then extended beyond the lowest
Landau level (LLL) \cite{Li}.

In this paper we calculate the effect of fluctuations on the magnetic field
distribution, structure function of the vortex lattice and compare with
existing MC results. Fluctuations cause the spread of the peaks in
diffraction pattern in a very specific way, while height of the peaks is
slightly corrected. Effects of fluctuation and disorder on magnetization and
specific heat are computed. The paper is organized as follows. In section II
the model and the fluctuation spectrum approximation are briefly reviewed.
In section III the \ calculation of the structure function is presented.
Section IV contains analysis of the result, comparison with MC simulation
and some generalizations. In section V the distribution of magnetic field is
calculated, while effects of weak disorder on magnetization and specific
heat are treated in section VI. Summary is given in section VII.

\section{Model, mean field solution and the perturbation theory}

Our starting point is the GL free energy: 
\begin{equation}
F=\int d^{3}x\frac{{\hbar }^{2}}{2m_{ab}}\left| \left( {\bf \nabla }-\frac{%
ie^{\ast }}{\hbar c}{\bf A}\right) \psi \right| ^{2}+\frac{{\hbar }^{2}}{%
2m_{c}}|\partial _{z}\psi |^{2}+a|\psi |^{2}+\frac{b^{\prime }}{2}|\psi |^{4}
\label{energy}
\end{equation}
Here ${\bf A}=(-By,0)$ describes a nonfluctuating constant magnetic field.
For strongly type II superconductors ($\kappa \sim 100$) far from $H_{c1}$%
(this is the range of interest in this paper) magnetic field is homogeneous
to a high degree due to superposition from many vortices. For simplicity we
assume $a=\alpha (1-t)$, $t\equiv T/T_{c}$ although this dependence can be
easily modified to better describe the experimental coherence length.

Throughout most of the paper will use the following units. Unit of length is 
$\xi =\sqrt{{\hbar }^{2}/\left( 2m_{ab}\alpha T_{c}\right) }$ and unit of
magnetic field is $H_{c2}$, so that dimensionless magnetic field is $b\equiv
B/H_{c2}$. The dimensionless free energy in these units is (the order
parameter field is rescaled as $\psi ^{2}\rightarrow \frac{2\alpha T_{c}}{%
b^{\prime }}\psi ^{2}$): 
\begin{equation}
\frac{F}{T}=\frac{1}{\omega }\int d^{3}x\left[ \frac{1}{2}|D\psi |^{2}+\frac{%
1}{2}|\partial _{z}\psi |^{2}-\frac{1-t}{2}|\psi |^{2}+\frac{1}{2}|\psi |^{4}%
\right] ,  \label{energ1}
\end{equation}
The dimensionless coefficient is 
\begin{equation}
\omega =\sqrt{2Gi}\pi ^{2}t.  \label{omega}
\end{equation}
where the Ginzburg number is defined by $Gi\equiv \frac{1}{2}(\frac{32\pi
e^{2}\kappa ^{2}\xi T_{c}\gamma ^{1/2}}{c^{2}h^{2}})^{2}$ and $\gamma \equiv
m_{c}/m_{ab}$ is an anisotropy parameter. This coefficient determines the
strength of fluctuations, but is irrelevant as far as mean field solutions
are concerned.

The second expansion parameter is (see \cite{Ikeda,Li} for details): 
\begin{equation}
a_{h}\equiv \frac{1-t-b}{2}.  \label{ah}
\end{equation}
If $a_{h}$ is sufficiently small GL equations\ can be solved perturbatively: 
\begin{equation}
\psi =\Phi =(a_{h})^{1/2}\left[ \Phi _{0}+a_{h}\Phi _{1}+...\right]
\label{phi}
\end{equation}
It is convenient to represent $\Phi _{0},\Phi _{1},...$ in the basis of
eigenfunctions of operator ${\cal H}\equiv $ $\frac{1}{2}(-D^{2}-b)$, ${\cal %
H}\varphi ^{n}=nb\varphi ^{n}$, normalized to unit ''Cooper pairs density'' $%
<|\varphi ^{n}|^{2}>\equiv \int_{cell}d^{2}x|\varphi ^{n}|^{2}\frac{b}{2\pi }%
=1$, where ''cell'' is a primitive \ \ \ cell of the vortex lattice.
Assuming hexagonal lattice symmetry one explicitly has: 
\begin{eqnarray}
\varphi ^{n} &=&\sqrt{\frac{2\pi }{\sqrt{\pi }2^{n}n!a}}\sum\limits_{l=-%
\infty }^{\infty }H_{n}(y\sqrt{b}-\frac{2\pi }{a}l) \\
&&\times \exp \left\{ i\left[ \frac{\pi l(l-1)}{2}+\frac{2\pi \sqrt{b}}{a}lx%
\right] -\frac{1}{2}(y\sqrt{b}-\frac{2\pi }{a}l)^{2}\right\}  \nonumber
\end{eqnarray}
where $\frac{a}{\sqrt{b}}=\sqrt{\frac{4\pi }{\sqrt{3}b}}$ is the lattice
spacing. One finds 
\begin{equation}
\Phi _{0}=\frac{1}{\sqrt{\beta _{A}}}\varphi .
\end{equation}
To order $a_{h}^{i}$, we expand 
\begin{equation}
\Phi _{i}=g_{i}\varphi +\sum_{n=1}^{\infty }g_{i}^{n}\varphi ^{n}.
\end{equation}
These coefficients can be found in \cite{Li}.

To find an excitation spectrum one expands free energy functional around the
solution. The fluctuating order parameter field $\psi $ is divided into a
nonfluctuating (mean field) part and a small fluctuation 
\begin{equation}
\psi (x)=\Phi (x)+\chi (x).
\end{equation}
Field $\chi $ can be expanded in a basis of quasimomentum eigenfunctions:

\begin{eqnarray}
\varphi _{{\bf k}}^{n} &=&\sqrt{\frac{2\pi }{\sqrt{\pi }2^{n}n!a}}%
\sum\limits_{l=-\infty }^{\infty }H_{n}(y\sqrt{b}+\frac{k_{x}}{\sqrt{b}}-%
\frac{2\pi }{a}l) \\
&&\times \exp \left\{ i\left[ \frac{\pi l(l-1)}{2}+\frac{2\pi (\sqrt{b}x-%
\frac{k_{y}}{\sqrt{b}})}{a}l-xk_{x}\right] -\frac{1}{2}(y\sqrt{b}+\frac{k_{x}%
}{\sqrt{b}}-\frac{2\pi }{a}l)^{2}\right\}  \nonumber
\end{eqnarray}
Instead of complex field $\chi _{k}^{n}$we will use two ''real'' fields $%
O_{k}^{n}$ and $A_{k}^{n}$ satisfying $O_{k}^{n}=O_{-k}^{\ast n}$,$%
A_{k}^{n}=A_{-k}^{\ast n}:$%
\begin{eqnarray}
\chi (x) &=&\frac{1}{\sqrt{2}}\int_{k}d_{k}\frac{e^{-ik_{3}x_{3}}}{\sqrt{%
2\pi }}\sum_{n=0}^{\infty }\frac{\varphi _{{\bf k}}^{n}(x)}{\left( \sqrt{%
2\pi }\right) ^{2}}\left( O_{k}^{n}+iA_{k}^{n}\right)  \label{expan} \\
\chi ^{\ast }(x) &=&\frac{1}{\sqrt{2}}\int_{k}d_{k}^{\ast }\frac{%
e^{ik_{3}x_{3}}}{\sqrt{2\pi }}\sum_{n=0}^{\infty }\frac{\varphi _{{\bf k}%
}^{\ast n}(x)}{\left( \sqrt{2\pi }\right) ^{2}}\left(
O_{-k}^{n}-iA_{-k}^{n}\right)  \nonumber
\end{eqnarray}
where $d_{k}=\exp [-i\theta _{k}/2]$ where $\gamma _{k}=|\gamma _{k}|\exp
[i\theta _{k}]$. Within the LLL, the eigenstates are $A_{k}^{n},O_{k}^{n},$
while the eigenvalues (in two dimensions; in three dimensions simply plus $%
\frac{k_{3}^{2}}{2}$) are 
\begin{eqnarray}
\epsilon _{A} &=&a_{h}\epsilon _{A}^{1}=a_{h}\left( -1+\frac{2}{\beta }\beta
_{k}-\frac{1}{\beta }|\gamma _{k}|\right)  \label{elspec} \\
\epsilon _{O} &=&a_{h}\epsilon _{O}^{1}=a_{h}\left( -1+\frac{2}{\beta }\beta
_{k}+\frac{1}{\beta }|\gamma _{k}|\right) .  \nonumber
\end{eqnarray}
where $\epsilon _{A},\epsilon _{O}$ are dependent on two dimensional vector $%
{\bf k}$. Higher order corrections and higher Landau levels eigenstates and
eigenvalues can be found in \cite{Li}. With spectrum of excitations and
expansion of solutions of GL equations in $a_{h}$ one can start the
calculation of correlators to any order in $\omega .$

\section{Structure function of the vortex lattice}

In this section the structure function is calculated to order $\omega $
(harmonic approximation) within the LLL, namely neglecting higher $a_{h}$
corrections. We discuss these corrections in the next section. First we
calculate the density correlator defined by 
\begin{equation}
\widetilde{S}({\bf z},z_{3})=\left\langle \rho ({\bf x},x_{3})\rho ({\bf x}+%
{\bf z},x_{3}+z_{3})\right\rangle _{{\bf x}}=\left\langle \rho (x)\rho
(y)\right\rangle _{{\bf x}}
\end{equation}
where $<>_{{\bf x}}$indicates average over ${\bf x}$ (which means here over
the unit cell) and $\rho \equiv \left| \psi \right| ^{2}$. The correlator is
calculated using Wick expansion: 
\begin{equation}
\widetilde{S}=\widetilde{S}_{mf}+\omega \widetilde{S}_{fluct}.
\end{equation}
The first term is the mean field part, while the second term is the
fluctuation part.\ The mean field part is simply 
\begin{equation}
\widetilde{S}_{mf}=\left\langle |\Phi (x)|^{2}|\Phi (y)|^{2}\right\rangle _{%
{\bf x}}.
\end{equation}
The structure function is the Fourier transform $S({\bf q},0)=\int d{\bf z}%
e^{i{\bf q}\cdot {\bf z}}\widetilde{S}({\bf z},z_{3}=0)$. Within the LLL, $%
\Phi (x)=\left( \frac{a_{h}}{\beta _{A}}\right) ^{1/2}\varphi (x)$ and the
mean field part of the structure function becomes: 
\begin{eqnarray}
S_{mf}({\bf q},0) &\equiv &\int d{\bf z}e^{i{\bf q}\cdot {\bf z}}<|\Phi
(x)|^{2}|\Phi (y)|^{2}>_{{\bf x}}  \nonumber \\
&=&\left( \frac{a_{h}}{\beta _{A}}\right) ^{2}\frac{b}{2\pi }%
\int_{cell}|\varphi (x)|^{2}e^{-i{\bf q}\cdot {\bf x}}\int_{{\bf z}}|\varphi
(z)|^{2}e^{i{\bf q}\cdot {\bf z}} \\
&=&\left( \frac{a_{h}}{\beta _{A}}\right) ^{2}4\pi ^{2}\delta _{n}({\bf q}%
)\exp \left[ -\frac{{\bf q}^{2}}{2b}\right] ,  \nonumber
\end{eqnarray}
where we made use of formulas ..and function $\delta _{n}({\bf q})$ defined
in Appendix. This is just sum of delta functions of various heights at
reciprocal lattice points.

\ The fluctuation part contains four terms (diagrams) $\widetilde{S}_{1},...%
\widetilde{S}_{4}$. The first term is 
\begin{eqnarray}
\widetilde{S}_{1}({\bf z},z_{3}) &=&\frac{1}{4\pi \cdot 2\pi \omega }%
\left\langle \Phi (x)\Phi (y)\int_{k,l}d_{k}^{\ast }d_{l}^{\ast }\varphi _{%
{\bf k}}^{\ast n}(x)\varphi _{{\bf l}}^{\ast n}(y)\right\rangle _{{\bf x}} 
\nonumber \\
&&\times (<O_{k}^{\ast n}O_{l}^{\ast n}-A_{k}^{\ast n}A_{l}^{\ast
n}>)e^{ik_{3}(y-x)_{3}}+c.c.
\end{eqnarray}
$<O_{k}^{n}O_{l}^{n}>$ and $<A_{k}^{n}A_{l}^{n}>$ are propagators: 
\begin{eqnarray}
&<&O_{k}^{n}O_{l}^{n}>=\frac{\omega }{\epsilon _{O}^{n}({\bf k})+\frac{%
k_{3}^{2}}{2}}\delta ({\bf k}+{\bf l})  \label{prog} \\
&<&A_{k}^{n}A_{l}^{n}>=\frac{\omega }{\epsilon _{A}^{n}({\bf k})+\frac{%
k_{3}^{2}}{2}}\delta ({\bf k}+{\bf l})  \nonumber
\end{eqnarray}
To calculate structure functions we will need only the $\ z_{3}=0$
correlator: 
\begin{eqnarray}
\widetilde{S}_{1}({\bf z},0) &=&\frac{1}{4\left( 2\pi \right) ^{2}}%
\left\langle \Phi (x)\Phi (y)\int_{{\bf k}}\left( d_{k}^{\ast }\right)
^{2}\varphi _{{\bf k}}^{\ast n}(x)\varphi _{-{\bf k}}^{\ast
n}(y)\right\rangle _{{\bf x}}  \nonumber \\
&&\times \left[ \sqrt{\frac{2}{\epsilon _{O}^{n}({\bf k})}}-\sqrt{\frac{2}{%
\epsilon _{A}^{n}({\bf k})}}\right] +c.c.
\end{eqnarray}
Within the LLL approximation it simplified to: 
\begin{eqnarray}
\widetilde{S}_{1}({\bf z},0) &=&\frac{1}{4\left( 2\pi \right) ^{2}}\frac{%
a_{h}}{\beta _{A}}\left\langle \varphi (x)\varphi (y)\int_{{\bf k}}\left(
d_{k}^{\ast }\right) ^{2}\varphi _{{\bf k}}^{\ast }(x)\varphi _{-{\bf k}%
}^{\ast }(y)\right\rangle _{{\bf x}}  \nonumber \\
&&\times \left[ \sqrt{\frac{2}{\epsilon _{O}({\bf k})}}-\sqrt{\frac{2}{%
\epsilon _{A}({\bf k})}}\right] +c.c.
\end{eqnarray}

The first fluctuation correction term to structure function can be evaluated
as follows: 
\begin{eqnarray}
S_{1}({\bf q},0) &=&\frac{1}{4\left( 2\pi \right) ^{2}}\frac{a_{h}}{\beta
_{A}}\int_{{\bf k}}\frac{b}{2\pi }\int_{cell}\varphi (x)\varphi _{{\bf k}%
}^{\ast }(x)e^{-i{\bf q}\cdot {\bf x}}\int_{{\bf z}}\varphi (z)\varphi _{-%
{\bf k}}^{\ast }(z)e^{i{\bf q}\cdot {\bf z}}  \nonumber \\
&&\times \left( d_{k}^{\ast }\right) ^{2}\left[ \sqrt{\frac{2}{\epsilon _{O}(%
{\bf k})}}-\sqrt{\frac{2}{\epsilon _{A}({\bf k})}}\right] +c.c. \\
&=&\frac{a_{h}}{2\beta _{A}}\cos \left( \frac{k_{x}k_{y}+{\bf k}\times {\bf Q%
}}{b}+\theta _{k}\right) \exp \left[ -\frac{{\bf q}^{2}}{2b}\right] \left[ 
\sqrt{\frac{2}{\epsilon _{O}({\bf k})}}-\sqrt{\frac{2}{\epsilon _{A}({\bf k})%
}}\right]  \nonumber
\end{eqnarray}
where formulas of Appendix were used.\ ${\bf Q}$ is the integer part of $%
{\bf q}$, \ ${\bf k}$ is the fractional part of ${\bf q}$: ${\bf q}={\bf k}%
+n_{1}\widetilde{{\bf d}}_{1}+n_{2}\widetilde{{\bf d}}_{2}={\bf k}+$\ ${\bf Q%
}$ (see Appendix for the definitions of $\widetilde{{\bf d}}_{1},\widetilde{%
{\bf d}}_{2}$). The second fluctuation correction term is 
\begin{eqnarray}
\widetilde{S}_{2}({\bf z},z_{3}) &=&\frac{1}{4\pi \cdot \left( 2\pi \right)
^{2}\omega }\left\langle \Phi (x)\Phi ^{\ast }(y)\int_{k,l}d_{k}^{\ast
}d_{l}\varphi _{{\bf k}}^{\ast n}(x)\varphi _{{\bf l}}^{n}(y)\right\rangle _{%
{\bf x}}  \nonumber \\
& &\times ( <O_{k}^{\ast n}O_{l}^{n}+A_{k}^{\ast n}A_{l}^{\ast
n}>)e^{ik_{3}(y-x)_{3}}+c.c.
\end{eqnarray}
$\widetilde{S}_{2}({\bf z},z_{3}=0)$ is equal to (in the LLL approximation): 
\begin{eqnarray}
\widetilde{S}_{2}({\bf z},0) &=&\frac{1}{4\left( 2\pi \right) ^{2}}\frac{%
a_{h}}{\beta _{A}}\left\langle \varphi (x)\varphi ^{\ast }(y)\int_{{\bf k}%
}\varphi _{{\bf k}}^{\ast }(x)\varphi _{{\bf k}}(y)\right\rangle _{{\bf x}} 
\nonumber \\
&&\times \left[ \sqrt{\frac{2}{\epsilon _{O}({\bf k})}}+\sqrt{\frac{2}{%
\epsilon _{A}({\bf k})}}\right] +c.c.
\end{eqnarray}
and 
\begin{eqnarray}
S_{2}({\bf q},0) &=&\frac{1}{4\left( 2\pi \right) ^{2}}\frac{a_{h}}{\beta
_{A}}\int_{{\bf k}}\frac{b}{2\pi }\int_{cell}\varphi (x)\varphi _{{\bf k}%
}^{\ast }(x)e^{-i{\bf q}\cdot {\bf x}}\int_{{\bf z}}\varphi ^{\ast
}(z)\varphi _{{\bf k}}(z)e^{i{\bf q}\cdot {\bf z}}  \nonumber \\
&&\times \left( d_{k}^{\ast }\right) ^{2}\left[ \sqrt{\frac{2}{\epsilon _{O}(%
{\bf k})}}-\sqrt{\frac{2}{\epsilon _{A}({\bf k})}}\right] +c.c. \\
&=&\frac{a_{h}}{2\beta _{A}}\exp \left[ -\frac{{\bf q}^{2}}{2b}\right] \left[
\sqrt{\frac{2}{\epsilon _{O}({\bf k})}}+\sqrt{\frac{2}{\epsilon _{A}({\bf k})%
}}\right]  \nonumber
\end{eqnarray}

The third term is 
\begin{eqnarray}
\widetilde{S}_{3}({\bf z},z_{3}) &=&\frac{1}{4\pi \cdot \left( 2\pi \right)
^{2}\omega }\left\langle |\Phi (x)|^{2}\int_{k,l}d_{k}d_{l}^{\ast }\varphi _{%
{\bf k}}^{n}(y)\varphi _{{\bf l}}^{\ast n}(y)\right\rangle _{{\bf x}} 
\nonumber \\
&&\times (<O_{k}^{\ast n}O_{l}^{\ast n}+A_{k}^{\ast n}A_{l}^{\ast n}>)+{\bf x%
}\longleftrightarrow {\bf y}
\end{eqnarray}
and within the LLL at $z_{3}=0$ is equal to: 
\begin{eqnarray}
\widetilde{S}_{3}({\bf z},0) &=&\frac{1}{4\left( 2\pi \right) ^{2}}%
\left\langle |\Phi (x)|^{2}\int_{{\bf k}}\varphi _{{\bf k}}(y)\varphi _{{\bf %
k}}^{\ast }(y)\right\rangle _{{\bf x}}  \nonumber \\
&&\times \left[ \sqrt{\frac{2}{\epsilon _{O}({\bf k})}}+\sqrt{\frac{2}{%
\epsilon _{A}({\bf k})}}\right] +({\bf x}\longleftrightarrow {\bf y}).
\end{eqnarray}
Consequently the correction to the structure function is: 
\begin{eqnarray}
S_{3}({\bf q},0) &=&\frac{a_{h}}{4\beta _{A}}\int_{{\bf k}}\frac{b}{2\pi }%
\int_{cell}|\varphi (x)|^{2}e^{-i{\bf q}\cdot {\bf x}}\int_{{\bf z}}|\varphi
_{{\bf k}}(z)|^{2}e^{i{\bf q}\cdot {\bf z}}\times \left[ \sqrt{\frac{2}{%
\epsilon _{O}({\bf k})}}+\sqrt{\frac{2}{\epsilon _{A}({\bf k})}}\right] +(%
{\bf q}\rightarrow -{\bf q})  \nonumber \\
&=&\frac{a_{h}}{2\beta _{A}}\delta _{n}({\bf q})\exp \left[ -\frac{{\bf q}%
^{2}}{2b}\right] \int_{{\bf k}}\cos (\frac{{\bf k}\times {\bf Q}}{b})\left[ 
\sqrt{\frac{2}{\epsilon _{O}({\bf k})}}+\sqrt{\frac{2}{\epsilon _{A}({\bf k})%
}}\right]
\end{eqnarray}
The final term is from the vacuum renormalization contribution. The shift $v$
in $\psi (x)=v\phi (x)+\chi (x)$ is renormalized, that is, to one loop
order, $v^{2}=v_{o}^{2}+\omega v_{1}^{2},$ where $v_{o}^{2}=\frac{a_{h}}{%
\beta _{A}}$. One can find $v_{1}^{2}$ by minimizing the effective one loop
free energy:$\ $ 
\begin{equation}
\frac{1}{\omega }\left[ -a_{h}v^{2}+\frac{1}{2}\beta v^{4}\right] +\frac{1}{2%
}\left\{ Tr\ln \left[ 2\epsilon _{O}({\bf k})+k_{z}^{2}\right] +Tr\ln \left[
2\epsilon _{A}({\bf k})+k_{z}^{2}\right] \right\} ,
\end{equation}
where we write 
\begin{eqnarray}
\epsilon _{A}({\bf k}) &=&-a_{h}+2v^{2}\beta _{k}-v^{2}|\gamma _{k}| \\
\epsilon _{O}({\bf k}) &=&-a_{h}+2v^{2}\beta _{k}+v^{2}|\gamma _{k}|. 
\nonumber
\end{eqnarray}
Straightforward calculation gives: 
\begin{equation}
v_{1}^{2}=-\frac{1}{16\pi ^{2}}\int_{{\bf k}}\left[ \left( \frac{2\beta
_{k}+|\gamma _{k}|}{\beta }\right) \sqrt{\frac{2}{\epsilon _{O}({\bf k})}}%
+\left( \frac{2\beta _{k}-|\gamma _{k}|}{\beta }\right) \sqrt{\frac{2}{%
\epsilon _{A}({\bf k})}}\right]  \label{nu1}
\end{equation}

The last contribution to the one loop correction to the correlator is
therefore: 
\begin{eqnarray}
S_{4}({\bf z},z_{3}) &=&2\frac{a_{h}}{\beta _{A}}\left\langle |\varphi
(x)|^{2}|\varphi (y)|^{2}\right\rangle _{{\bf x}}\left( v_{1}\right) ^{2} 
\nonumber \\
S_{4}({\bf q},0) &=&\frac{2a_{h}}{\beta _{A}}\frac{b}{2\pi }%
\int_{cell}|\varphi (x)|^{2}e^{-i{\bf q}\cdot {\bf x}}\int_{{\bf z}}|\varphi
(z)|^{2}e^{i{\bf q}\cdot {\bf z}}v_{1}^{2} \\
&=&-\frac{a_{h}}{2\beta _{A}}\delta _{n}({\bf q})\exp \left[ -\frac{{\bf q}%
^{2}}{2b}\right] \int_{{\bf k}}{\Huge [}\left( \frac{2\beta _{k}+|\gamma
_{k}|}{\beta }\right) \sqrt{\frac{2}{\epsilon _{O}(k)}}  \nonumber \\
&&+\left( \frac{2\beta _{k}-|\gamma _{k}|}{\beta }\right) \sqrt{\frac{2}{%
\epsilon _{A}(k)}}{\Huge ]}.  \nonumber
\end{eqnarray}
The sum of all the four terms can be cast in the following form: 
\begin{eqnarray}
S({\bf q},0) &=&\left( \frac{a_{h}}{\beta _{A}}\right) ^{2}4\pi ^{2}\delta
_{n}({\bf q})\exp \left[ -\frac{{\bf q}^{2}}{2b}\right] +\frac{\omega }{2}%
\frac{a_{h}^{1/2}}{\beta _{A}}\exp \left[ -\frac{{\bf q}^{2}}{2b}\right] 
\nonumber \\
&&\times \left[ f_{1}({\bf q})+\delta _{n}({\bf q})f_{2}({\bf Q})+\delta
_{n}({\bf q})f_{3}\right]  \nonumber \\
f_{1}({\bf q}) &=&\left[ 1+\cos \left( \frac{k_{x}k_{y}+{\bf k}\times {\bf Q}%
}{b}+\theta _{k}\right) \right] \sqrt{\frac{2}{\epsilon _{O}^{1}({\bf k})}}
\label{fde} \\
&&+\left[ 1-\cos (\frac{k_{x}k_{y}+{\bf k}\times {\bf Q}}{b}+\theta _{k})%
\right] \sqrt{\frac{2}{\epsilon _{A}^{1}({\bf k})}}  \nonumber \\
f_{2}({\bf Q}) &=&\int_{{\bf k}}\left[ -1+\cos \left( \frac{{\bf k}\times 
{\bf Q}}{b}\right) \right] \left[ \sqrt{\frac{2}{\epsilon _{O}^{1}({\bf k})}}%
+\sqrt{\frac{2}{\epsilon _{A}^{1}({\bf k})}}\right]  \nonumber \\
f_{3} &=&-\int_{{\bf k}}\left[ \sqrt{2\epsilon _{O}^{1}({\bf k})}+\sqrt{%
2\epsilon _{A}^{1}({\bf k})}\right] =-28.5275b.  \nonumber
\end{eqnarray}
The reason we write the sum in such a form will be explained in the next
section.

\section{Analysis of the result, comparison with MC simulations}

Although all of the four terms $S_{1}$, $S_{2},$ $S_{3}$ and $S_{4}$ are
divergent as any of the peaks is approached, ${\bf k}\rightarrow {\bf 0}$,
the sums $S_{1},S_{2}$ and $S_{3},S_{4}$ are not. We start with the first
two: 
\begin{equation}
S_{1}({\bf q},0)+S_{2}({\bf q},0)=\frac{\omega }{2}\frac{a_{h}}{\beta _{A}}%
\exp \left[ -\frac{{\bf q}^{2}}{2b}\right] f_{1}({\bf q}),
\end{equation}
where $f_{1}({\bf q})$ defined in eq.(\ref{fde}) contains a function $\frac{1%
}{b}(k_{x}k_{y}+{\bf k}\times {\bf Q)}+\theta _{k}$ . When ${\bf k}%
\rightarrow {\bf 0}$ it can be shown that $\frac{k_{x}k_{y}}{b}+\theta
_{k}=O\left( {\bf k}^{2\cdot 2}\right) $ , thus $\frac{1}{b}(k_{x}k_{y}+{\bf %
k}\times {\bf Q)}+\theta _{k}$ $\rightarrow {\bf k}\times {\bf Q}$ , and \ $%
1-\cos (\frac{k_{x}k_{y}+{\bf k}\times {\bf Q}}{b}+\theta _{k})\rightarrow
\left( {\bf k}\times {\bf Q}\right) ^{2}$. Hence it will cancel the $1/k^{2}$
singularity coming from $\sqrt{\frac{1}{\epsilon _{A}^{1}(k)}}$. Thus $f_{1}(%
{\bf q})$ approaches $const.+const.\cdot \frac{\left( {\bf k}\times {\bf Q}%
\right) ^{2}}{k^{2}}$ when ${\bf Q}\neq {\bf 0}$, and approaches $%
const.+const.\cdot k^{6}$ when ${\bf Q}={\bf 0}$.

Similarly the sum of $S_{4}({\bf q},0)$ and $S_{3}({\bf q},0)$ is not
divergent, although separately they are. Their sum is. 
\begin{equation}
S_{3}({\bf q},0)+S_{4}({\bf q},0)=\frac{\omega }{2}\frac{a_{h}^{1/2}}{\beta
_{A}}\delta _{n}({\bf q})\exp \left[ -\frac{{\bf q}^{2}}{2b}\right] \left[
f_{2}({\bf Q})+f_{3}\right]
\end{equation}
Now we compare our results with numerical simulation of the LLL system in 
\cite{Sasik}. The general shape of the structure function in the vicinity of
a peak \ (see Fig. 1b) and the data near the origin according to a MC
simulation of the same system within the same LLL approximation in ref. \cite
{Sasik} (Fig. 1a) are qualitatively same pattern. It is easier to compare
using rescaled quasimomenta: ${\bf q}\rightarrow {\bf q}\sqrt{b},{\bf k}%
\rightarrow {\bf k}\sqrt{b}$. We get 
\begin{eqnarray}
S({\bf q},0) &=&\left( \frac{a_{h}}{\beta _{A}}\right) ^{2}4\pi ^{2}\frac{%
\delta _{n}({\bf q})}{b}\exp \left[ -\frac{{\bf q}^{2}}{2}\right] \\
&&+\frac{\omega }{2}\frac{a_{h}^{1/2}}{\beta _{A}}\exp \left[ -\frac{{\bf q}%
^{2}}{2}\right] \left[ f_{1}({\bf q})+\delta _{n}({\bf q})f_{2}({\bf Q}%
)+\delta _{n}({\bf q})f_{3}\right]  \nonumber
\end{eqnarray}
where $f_{1}({\bf q}),f_{2}({\bf Q})$ and $f_{3}$ are defined in eq.(\ref
{fde}), but with $b=1$ (for example, $f_{3}=-28.5275$) and the region of
integration in the formula rescaled to the \ cell with $\widetilde{{\bf d}}%
_{1},\widetilde{{\bf d}}_{2}$ \ being the reciprocal lattice basis vectors 
\[
\widetilde{{\bf d}}_{1}=\frac{2\pi }{a}\left( 1,-\frac{1}{\sqrt{3}}\right)
;\;\widetilde{{\bf d}}_{2}=\left( 0,\frac{4\pi }{a\sqrt{3}}\right) 
\]
Furthermore we define $s({\bf q})$ which is used also in \cite{Sasik}: 
\begin{eqnarray}
S({\bf q},0) &=&\left( \frac{a_{h}}{\beta _{A}}\right) ^{2}\frac{4\pi ^{2}}{b%
}\exp \left[ -\frac{{\bf q}^{2}}{2}\right] s({\bf q}) \\
s({\bf q}) &\equiv &\delta _{n}({\bf q})+\frac{\beta _{A}b\omega \left(
a_{h}\right) ^{-3/2}}{8\pi ^{2}}\left[ f_{1}({\bf q})+\delta _{n}({\bf q}%
)f_{2}({\bf Q})+\delta _{n}({\bf q})f_{3}\right]  \nonumber
\end{eqnarray}
For reciprocal lattice vectors close to origin the value of $f_{2}({\bf Q})$
are:

\begin{center}
{\bf Table 1.}

Values of $f_{2}({\bf Q})$ \ with small $n_{1,}n_{2}$.

\begin{tabular}{|c|c|c|c|c|c|c|c|}
\hline
$n_{1,}n_{2}$ & $0,1$ & $1,0$ & $1,1$ & $1,2$ & $0,2$ & $2,2$ & $1,3$ \\ 
\hline
$f_{2}({\bf Q})$ & $-46.19$ & $-46.19$ & $-46.19$ & $-63.165$ & $-64.35$ & $%
-64.35$ & $-73.825$ \\ \hline
\end{tabular}
\end{center}

\bigskip

In ref. \cite{Sasik}, the material parameters describe YBCO: $T_{c}=93$~K, $%
dH_{c2}(T)/dT=-1.8\times 10^{4}$~Oe/K, $\gamma =5$, and $\kappa =52$. At $%
T=82.8$K, $H=50$ kOe. This leads to following dimensionless parameters $%
Gi=2.04\times 10^{-5}$, $\omega =0.056$, $a_{h}=0.039904$. However as
discussed in \cite{Li} effective expansion parameters are $\frac{a_{h}}{6b}%
=0.\,222\,68$ and $\frac{\omega \left( a_{h}\right) ^{-3/2}}{2\sqrt{2}\pi }%
=0.79328$, both less than one. $\frac{a_{h}}{6b}$ is the parameter for the
expansion of the classical solution. The factor $6$ comes from a fact that
due to hexagonal, thus only 6th, 12th, {\it etc}. Landau levels appears in
perturbation expansion. $\frac{\omega \left( a_{h}\right) ^{-3/2}}{2\sqrt{2}%
\pi }$ is the parameter for the fluctuations (loop) expansion and is just
barely less than one here. It justifies the quantum correction of the
formula using perturbation expansion. The numerical factor in front of the
fluctuation correction in this case is 
\begin{equation}
c_{1}=\frac{\beta _{A}b\omega \left( a_{h}\right) ^{-3/2}}{8\pi ^{2}}%
=3.\,093\,2\times 10^{-3}
\end{equation}
In a finite size sample, $\delta _{n}({\bf q})$ is equal to $\frac{L_{x}L_{y}%
}{\left( 2\pi \right) ^{2}}$ when ${\bf q}$ lies on the reciprocal lattice $%
m_{1}\widetilde{{\bf d}}_{1}+m_{2}\widetilde{{\bf d}}_{2}$, otherwise it is
zero. Because $L_{x}L_{y}=N_{\phi }2\pi $ ($N_{\phi }$ is the number of
vortices of order $100$ only in the MC simulation), thus it is equal to $%
\frac{N_{\phi }}{2\pi }$. The normalized structure function ($s_{n}({\bf 0}%
)=1$, as it was used in \cite{Sasik}) is: 
\begin{equation}
s_{n}({\bf q})=\Delta ({\bf q})+\frac{1}{(1+c_{1}f_{3})}\left[ c_{2}f_{1}(%
{\bf q})+c_{1}\Delta ({\bf q})f_{2}({\bf Q})\right]  \label{ploting}
\end{equation}
and 
\begin{eqnarray}
c_{2} & &=c_{1}\frac{2\pi }{N_{\phi }}=1.\,943\,5\times 10^{-4}  \nonumber \\
& & 1+c_{1}f_{3} =.\,913\,15
\end{eqnarray}
The correction to the height of the peak at ${\bf Q}$, $\frac{c_{1}\Delta (%
{\bf q})}{(1+c_{1}f_{3})}f_{2}({\bf Q})$, is quite small. We find the height
of peak away from origin found in the MC simulation \cite{Sasik} are
typically smaller than ours, while around the peaks are larger than
analytical. It may be due to finite size effect or finite samplings of MC
calculation. In the MC calculation part of the peak might ''belong'' to a
neighboring pixel. We plot the correction to the non-peak region on Fig. 1b
and find that the theoretical prediction has roughly the same characteristic
saddle shape ''halos'' around the peaks as in ref. \cite{Sasik}, Fig. 1a on
which all the peaks were removed (so it is different from Fig.2a in \cite
{Sasik} on which only the central peak was removed).

We can extend our formula to higher orders which will include also the HLLs
(higher Landau levels). To next order of $a_{h}$, we should include $\Phi
_{1}$ in $\Phi $, $\Phi =(a_{h})^{1/2}\left[ \Phi _{0}+a_{h}\Phi _{1}\right] 
$, \ consider $\epsilon _{O}(k),\epsilon _{A}(k)$ to next order $a_{h}^{2}$,
and $\epsilon _{O}^{n}(k),\epsilon _{A}^{n}(k)$ to order $a_{h}$. It is
straightforward to do it.

\section{Fluctuation of magnetic field}

Another quantity which can be measured is the magnetic field distribution.
In addition to constant magnetic field background there are $1/\kappa $
magnetization corrections due to field produced by supercurrent. To leading
order in $a_{h}$ it is given by $m(x)\propto \frac{<\rho (x)>}{\kappa }$(for
example, see ref. \cite{Affleck}). \ $\left\langle \rho (x)\right\rangle $
can be calculated using the following equation, 
\begin{eqnarray}
\left\langle \rho (x)\right\rangle &=&\left\langle \left| \Phi (x)+\chi
(x)\right| ^{2}\right\rangle  \nonumber \\
&=&\left| \Phi (x)\right| ^{2}+\left\langle \Phi ^{\ast }(x)\chi
(x)\right\rangle +\left\langle \Phi (x)\chi ^{\ast }(x)\right\rangle
+\left\langle \chi (x)\chi ^{\ast }(x)\right\rangle  \label{rox} \\
&=&\frac{a_{h}}{\beta _{A}}\left| \phi (x)\right| ^{2}+\left\langle \chi
(x)\chi ^{\ast }(x)\right\rangle .  \nonumber
\end{eqnarray}
Using eq.(\ref{prog}) and eq. (\ref{expan}), and considering only $x_{3}=0$,
one obtains: 
\begin{eqnarray*}
\left\langle \chi (x)\chi ^{\ast }(x)\right\rangle &=&\frac{\omega }{16\pi
^{3}}\int_{k}\left| \varphi _{{\bf k}}(x)\right| ^{2}\left[ \frac{1}{%
\epsilon _{O}({\bf k})+\frac{k_{3}^{2}}{2}}+\frac{1}{\epsilon _{A}({\bf k})+%
\frac{k_{3}^{2}}{2}}\right] \\
&=&\frac{\omega }{16\pi ^{2}}\int_{{\bf k}}\left| \varphi _{{\bf k}%
}(x)\right| ^{2}\left[ \sqrt{\frac{2}{\epsilon _{O}({\bf k})}}+\sqrt{\frac{2%
}{\epsilon _{A}({\bf k})}}\right]
\end{eqnarray*}
However, as pointed out in Sec. III the coefficient $\nu $ in $\psi
(x)=v\phi (x)+\chi (x)$ is renormalized to one loop order, $%
v^{2}=v_{o}^{2}+\omega v_{1}^{2},$ with $v_{1}$ given in eq.(\ref{nu1}).
Thus we need to add a term, $\omega v_{1}^{2}\left[ \phi (x)\right] ^{2}$ to
eq.(\ref{rox}). 
\begin{eqnarray}
\left\langle \rho ({\bf x,0})\right\rangle &=&\frac{a_{h}}{\beta _{A}}\left|
\phi (x)\right| ^{2}+\frac{\omega }{16\pi ^{2}}\int_{{\bf k}}\left| \varphi
_{{\bf k}}(x)\right| ^{2}\left[ \sqrt{\frac{2}{\epsilon _{O}({\bf k})}}+%
\sqrt{\frac{2}{\epsilon _{A}({\bf k})}}\right] \\
&&-\;\frac{\omega \left| \phi (x)\right| ^{2}}{16\pi ^{2}}\int_{{\bf k}}%
\left[ \left( \frac{2\beta _{k}+|\gamma _{k}|}{\beta }\right) \sqrt{\frac{2}{%
\epsilon _{O}({\bf k})}}+\left( \frac{2\beta _{k}-|\gamma _{k}|}{\beta }%
\right) \sqrt{\frac{2}{\epsilon _{A}({\bf k})}}\right]  \nonumber
\end{eqnarray}
Its Fourier transform $\rho ({\bf q})\equiv \int d{\bf z}e^{i{\bf q}\cdot 
{\bf z}}\left\langle \rho ({\bf x,0})\right\rangle $ can be easily
calculated: 
\begin{eqnarray}
\rho ({\bf q}) &=&4\pi ^{2}\delta _{n}({\bf q})\exp \left[ -\frac{{\bf q}^{2}%
}{4b}+\frac{iq_{x}q_{y}}{2b}+\frac{\pi i}{2}(n_{1}^{2}-n_{1})\right] 
\nonumber \\
&&{\Huge \{}\frac{a_{h}}{\beta _{A}}+\frac{\omega }{16\pi ^{2}}\int_{{\bf k}}%
{\Huge [}\left( \exp \left( \frac{i{\bf k\times q}}{b}\right) -\left( \frac{%
2\beta _{k}+|\gamma _{k}|}{\beta }\right) \right) \sqrt{\frac{2}{\epsilon
_{O}({\bf k})}} \\
&&+\left( \exp \left( \frac{i{\bf k\times q}}{b}\right) -\left( \frac{2\beta
_{k}-|\gamma _{k}|}{\beta }\right) \right) \sqrt{\frac{2}{\epsilon _{A}({\bf %
k})}}{\Huge ]\}}  \nonumber
\end{eqnarray}
Performing integrals and rescaling the quasimomenta again, one obtains: 
\begin{eqnarray}
\rho ({\bf q}) &=&4\pi ^{2}\frac{\delta _{n}({\bf q})}{b}\exp \left[ -\frac{%
{\bf q}^{2}}{4}+\frac{iq_{x}q_{y}}{2}+\frac{\pi i}{2}(n_{1}^{2}-n_{1})\right]
\nonumber \\
&&{\Huge \{}\frac{a_{h}}{\beta _{A}}+\frac{\omega ba_{h}^{-1/2}}{16\pi ^{2}}%
\left[ -28.5275+f_{2}({\bf Q)}\right] {\Huge \}}
\end{eqnarray}
The function $f_{2}({\bf Q)}$ appeared in eq.(\ref{fde}).

\section{Disorder effect on magnetization and specific heat}

One can introduce weak disorder by adding a quadratic term in eq. (\ref
{energ1}) \cite{Blatter}, 
\begin{equation}
\Delta f\equiv \int d^{3}x\alpha (x)|\psi |^{2}.
\end{equation}
Loosely speaking it represents local variation of temperature. For pointlike
defects one can assume that the correlation of $\alpha (x)$ is $<<\alpha
(x)\alpha (y)>>=W\delta ({\bf x}-{\bf y}),<<\alpha (x)>>=0$. Before the
disorder average we calculate the free energy $-T\ln Z$ with 
\begin{equation}
Z=\int {\cal D}\psi ^{\ast }{\cal D}\psi \exp \left\{ -\frac{1}{\omega }%
\left[ f[\psi ^{\ast },\psi ]-\int d^{3}x\alpha (x)|\psi |^{2}\right]
\right\} .
\end{equation}
If $W$ is very small, we can calculate $Z$ by perturbation theory in $W$. To
the second order\ $Z$ is given as 
\begin{equation}
Z=Z_{0}\left[ 1-\frac{1}{\omega }\int_{x}\alpha (x)<\rho (x)>+\frac{1}{%
2\omega ^{2}}\int_{x}\int_{y}<\rho (x)\rho (y)>\alpha (x)\alpha (y)\right] ,
\end{equation}
where $Z_{0}$ is the free energy without disorder and it had been obtained
in ref. \cite{Li}. \ Thus the free energy with disorder \ is 
\begin{eqnarray}
F &=&-T\ln Z=F_{0}+\Delta F \\
&=&F_{0}+T\int_{x}\frac{\alpha (x)}{\omega }<\rho (x)>-\;\frac{T}{2\omega
^{2}}\int_{x}\int_{y}\left[ <\rho (x)\rho (y)>-<\rho (x)><\rho (y)>\right]
\alpha (x)\alpha (y),  \nonumber
\end{eqnarray}
where $F_{0}=-T\ln Z_{0}$. Averaging free energy over disorder one obtains: 
\begin{eqnarray}
F &=&F_{0}-\;\frac{TW}{2\omega ^{2}}\int_{x}\left[ <\rho (x)\rho (x)>-<\rho
(x)><\rho (x)>\right] \\
&=&F_{0}-\frac{TWV}{2\omega ^{2}}\omega \widetilde{S}_{fluct}(0).  \nonumber
\end{eqnarray}
From eq.(\ref{fde}), one finds that $\widetilde{S}_{fluct}(0)=-0.18619\frac{%
a_{h}^{1/2}b}{\beta _{A}}$ . Hence the energy density difference due to
disorder is ${\cal F}=\frac{\Delta F}{V}=-0.0931\frac{TWa_{h}^{1/2}b}{\omega
\beta _{A}}.$ Since $\omega =\sqrt{2Gi}\pi ^{2}t$ ${\cal F}=ca_{h}^{1/2}b$
with $c=-\frac{0.0931T_{c}W}{\sqrt{2Gi}\pi ^{2}\beta _{A}}$. The disorder
effect on magnetization and specific heat are 
\begin{eqnarray}
\Delta m &=&-\frac{\partial \Delta f}{\partial b}=-c\left( a_{h}^{1/2}-\frac{%
b}{4}a_{h}^{-1/2}\right) \\
\Delta c &=&-t\frac{\partial ^{2}}{\partial t^{2}}\Delta f=\frac{c}{16}%
tba_{h}^{-3/2}  \nonumber
\end{eqnarray}
respectively.

\section{Conclusions}

To conclude, we have calculated the effect of fluctuations on the structure
\ function of the vortex lattice and compared it to existing MC results. In
addition to renormalization of the height of the Bragg peaks, there appears
a characteristic saddle shape ''halos'' around the peaks as found in ref. 
\cite{Sasik}. The calculated fluctuations contribution to the magnetic field
can be more easily observed in low temperature strongly type II
superconductors. Finally the predicted dependence of magnetization and
specific heat on disorder via fluctuations also can be experimentally
studied.

Correlations in flux lattices can be experimentally measured using neutron
scattering as well as some other more exotic methods like muon spin
relaxation, electron tomography, scanning SQUID microscopy etc.\cite
{Gammel,Keimer,Maggio,Cubitt,Vu}. It would be interesting to detect the
effect of fluctuations given in the present paper directly from experiments
by subtracting the ''background'' of the well known mean field correlator.
The calculations show that infrared divergencies naively expected in all of
the physical quantities calculated above due to ''supersoft'' shear modes in
the large $\kappa $ limit cancel. This strengthens the view that the loop
expansion is a reliable theoretical tool to study the fluctuations effects
in vortex lattice below the melting point.

\acknowledgments
We are grateful to our colleagues A. Knigavko, B. Bako, V. Yang. One of us
(B.R.) is specially grateful to R. Sasik and D. Stroud for providing raw
numerical data which was essential for the present comparison with the MC
data. The work is part of the NCTS topical program on vortices in high Tc
and was supported by NSC of Taiwan. 
\newpage 
\appendix
\section{}
In this appendix, we present some basic formulas used in the calculations.
The basic matrix element is: 
\begin{equation}
\frac{b}{2\pi }\int_{cell}dx\varphi ({\bf x})\varphi _{{\bf k}}^{\ast }({\bf %
x})\exp [-i{\bf x}\cdot {\bf q}]=\Delta _{{\bf q,k}}\exp \left[ \frac{\pi i}{%
2}(n_{1}^{2}-n_{1})-\frac{{\bf q}^{2}}{4b}-\frac{iq_{x}q_{y}}{2b}+\frac{%
ik_{x}q_{y}}{b}\right]
\end{equation}
The Kronecker delta is defined by: 
\begin{equation}
\Delta _{{\bf q,k}}=\Delta ({\bf q}-{\bf k})=%
{1,\text{ if }{\bf q}={\bf k}+n_{1}\widetilde{{\bf d}}_{1}+n_{2}\widetilde{{\bf d}}_{2} \atopwithdelims\{. 0,\text{ otherwise}}%
.
\end{equation}
where integers $n_{1}=\frac{1}{2\pi }{\bf d}_{1}\cdot ({\bf q}-{\bf k})$ and 
$n_{2}=\frac{1}{2\pi }{\bf d}_{2}\cdot ({\bf q}-{\bf k})$. Here $\widetilde{%
{\bf d}}_{1},\widetilde{{\bf d}}_{2}$ are the reciprocal lattice basis
vectors 
\begin{equation}
\widetilde{{\bf d}}_{1}=\frac{2\pi \sqrt{b}}{a}\left( 1,-\frac{1}{\sqrt{3}}%
\right) ;\;\widetilde{{\bf d}}_{2}=\left( 0,\frac{4\pi \sqrt{b}}{a\sqrt{3}}%
\right) ,
\end{equation}
which are dual to $\ {\bf d}_{1}=(a/\sqrt{b},0),{\bf d}_{2}=\left( \frac{a}{2%
\sqrt{b}},\frac{a\sqrt{3}}{2\sqrt{b}}\right) $ and $a=\sqrt{\frac{4\pi }{%
\sqrt{3}}}$. Integrating over the sample area $A$ , one obtains: 
\begin{eqnarray}
\int_{A}dx\varphi ({\bf x})\varphi _{{\bf k}}^{\ast }({\bf x})\exp [-i{\bf x}%
\cdot {\bf q}] &=&4\pi ^{2}\delta _{n}({\bf q}-{\bf k})\exp \left[ \frac{\pi
i}{2}(n_{1}^{2}-n_{1})\right]  \nonumber \\
&&\times \exp \left[ -\frac{{\bf q}^{2}}{4b}-\frac{iq_{x}q_{y}}{2b}+\frac{%
ik_{x}q_{y}}{b}\right]
\end{eqnarray}
where $\delta _{n}({\bf q}-{\bf k})$ is defined as $\delta _{n}({\bf q}-{\bf %
k})=\sum_{m_{1,}m_{2}}\delta ({\bf q}-{\bf k}-m_{1}\widetilde{{\bf d}}%
_{1}-m_{2}\widetilde{{\bf d}}_{2})$.

\newpage

\begin{center}
{\Huge Figure captions}
\end{center}


{\LARGE Fig. 1a}

Structure factor from the MC simulation of ref.\cite{Sasik}. The peaks at
reciprocal lattice points are removed.

{\LARGE Fig. 1b}

Fluctuation correction to structure factor of the Abrikosov vortex lattice,
eq.(\ref{ploting}). The peaks at reciprocal lattice points are removed.

\end{document}